\documentclass[11pt,english]{article}

\renewcommand{\Re}{{\rm Re \, }}

\usepackage{euscript}
\usepackage{amssymb}
\usepackage{amsfonts}
\usepackage{amsbsy}
\usepackage{amsmath}
\usepackage{epsfig}
\usepackage{amsthm}
\usepackage{amscd}
\usepackage{amstext}

\usepackage{color}
\definecolor{darkgreen}{rgb}{0,0.5,0}
\definecolor{darkblue}{rgb}{0,0,0.6}
\definecolor{purple}{rgb}{0.4,.2,0.7}
\usepackage[colorlinks=true,citecolor=darkgreen,linkcolor=darkblue,urlcolor=purple]{hyperref}

\numberwithin{equation}{section}
\numberwithin{figure}{section}
\numberwithin{table}{section}

\def\IZ{{\mathbb Z}}

\begin{document}



\vskip 15mm 

$\;$\newline

\begin{center}
 {{\LARGE \textsc{Cosmic Clustering}} }  
\end{center}

\vskip5mm

\begin{center}
Dionysios Anninos$^1$ and Frederik Denef$^{2,3}$
\end{center}

\begin{center}

{
\it{$^1$ Stanford Institute for Theoretical Physics, Stanford University} 

\it{$^2$ Institute for Theoretical Physics, University of Leuven} }

\it{$^3$ Center for the Fundamental Laws of Nature, Harvard University} 

\vskip5mm

{\tt{danninos@stanford.edu,  frederik.denef@fys.kuleuven.be}}

\end{center}

\vskip15mm

\begin{abstract}

We show that the late time Hartle-Hawking wave function for a free massless scalar in a fixed de Sitter background encodes a sharp ultrametric structure for the standard Euclidean distance on the space of field configurations. This implies a hierarchical, tree-like organization of the state space, reflecting its genesis as a branched diffusion process. An equivalent mathematical structure organizes the state space of the Sherrington-Kirkpatrick model of a spin glass.

\end{abstract}

\newpage

\section{$\mathcal{I}^+$sosceles Inception}

As envisioned in \cite{anniehall}, an exponentially expanding universe, such as our own \cite{astro-ph/9805201,astro-ph/9812133,SLAC-PUB-2576,LEBEDEV-81-229}, continuously breaks apart: its wave function constantly separates in new branches in which super-horizon modes of the metric and other light fields take on definite expectation values. Cluster decomposition (factorization of correlators at large spatial separation) holds within each separate branch but not for the full quantum state obtained by superimposing them. This process can be effectively thought of as a branched diffusion process \cite{183959}, in which quantum fluctuations of light fields in a given branch get stretched out to super-horizon scales, freeze and classicalize, while falling out of causal contact with each other. This spawns an offspring of new branches, which in turn give birth to new branches, and so on.

The result --- pictorially at least --- is a tree-like structure on the state space of quantum fields on inflating spacetimes, and in particular on de Sitter space \cite{Denef:2011ee,diothesis,talks}. This should be distinguished from the tree-like causal structure of de Sitter space itself or models thereof \cite{gr-qc/0111048,arXiv:1110.0496}, where causally disconnected patches of space continuously spawn new causally disconnected patches of space. The two are of course related, as they both are produced by the peculiar generative dynamics of inflation.

The dynamical tree structure should be encoded somehow in the wave function of the universe at late times, and thus, assuming some version of a dS/CFT correspondence \cite{hep-th/0106113,hep-th/0106109,astro-ph/0210603,arXiv:1108.5735}, in the partition function of the putative dual field theory at ${\cal I}^+$. In particular it should emerge from the Hartle-Hawking wave function of for instance a massless free scalar on de Sitter space. At first sight this seems hard to imagine, as the Bunch-Davies/Hartle-Hawking vacuum wave function of a massless free scalar is an almost trivial Gaussian at all times \cite{134659,PRINT-83-0937 (CAMBRIDGE)}. Nevertheless, we will show in this paper that it emerges in a very sharp sense.

The approach we take is inspired by methods developed to analyze the state space of spin glasses \cite{spinglassbook}, in particular the remarkable Parisi solution \cite{LNF-79-31-P,parisi2,198330} of the Sherrington-Kirkpatrick model \cite{107911}. A crucial element in this solution was the presence of \emph{ultrametricity} \cite{198330,244332} in the state space: any three ``pure'' equilibrium states satisfy the property that their distances (measuring dissimilarity in local magnetization) form an isosceles triangle, with the unequal one the shortest of the three. Equivalently, these states can be organized as the leaves of a tree, with the distances corresponding to how far back on the tree one has to go to find a common branch. Yet another way of phrasing this is that the state space forms a hierarchy of nested clusters, with distances given by the size of the smallest common cluster. The Parisi solution involves an auxiliary flow equation reminiscent of a Hamilton-Jacobi equation in an expanding space or an RG equation (as reviewed in \cite{Denef:2011ee}), which generates the ultrametric structure. Although it is computationally intractable to enumerate the equilibrium states explicitly, it is possible to compute exact $n$-point \emph{probability distributions} of distances between equilibrium states, sampled from the Boltzmann-Gibbs measure. Such distance distributions serve as order parameters for the spin glass phase. Ultrametricity shows up in the fact that the three point distance distribution has support entirely on  isosceles triangles.

We will compute the one point and three point Euclidean field space distance distributions for a free massless scalar in dS$_{d+1}$ for arbitrary $d$, directly from the Hartle-Hawking wave function evaluated at late times, and without making any assumptions about effective classical behavior. In the Minkowski case, after fixing the zeromode, the distance distribution becomes a delta-function peaked at zero, since there is a unique vacuum, which satisfies cluster decomposition. But in de Sitter, as we will see, these distributions acquire a finite width, with a scale set by the de Sitter temperature. This confirms the Bunch-Davies vacuum does not satisfy cluster decomposition, but that it decomposes instead in an infinite set of branches which do. (See e.g.\ \cite{Denef:2011ee} for a more detailed explanation of the relation to the clustering property.) The distributions are always of Gumbel type. Gumbel distributions arise as extreme value distributions, i.e.\ distributions of maxima or minima of a set of random variables. They pop up in a wild variety of physical systems (see for example \cite{gumbel}). Interestingly, the same Gumbel distribution was found very recently in \cite{arXiv:1111.4195} as the distance distribution for a free massless scalar in dS$_2$, but for an a priori rather different distance measure, suggesting it is a universal feature.

More strikingly, we will demonstrate directly that the three point distance distribution exhibits  nontrivial ultrametricity: Given two distances $d_{12}$ and $d_{23}$ with $d_{12}$ less than $d_{23}$ by more than a few dS units, the conditional probability for the third distance $d_{31}$ peaks sharply on $d_{23}$, the longer one of the two. The dynamical origin of this hidden tree structure in the wave function at a fixed time is simple: The fields drift away from each other at a steady rate, so the distance between different states (labeled by large scale field vevs) at some fixed time slice is proportional to how far back in time one has to go to get to a common ancestor. What is remarkable is that this can be seen in a simple Gaussian wave functional.

\section{Defining Distance}

We begin with a brief discussion of the notion of distance between field configurations. We will restrict to spatially flat FLRW cosmologies.

\subsection{Geometry}

The $(d+1)$-dimensional FLRW geometry with scale factor $a(\eta)$ is given by:
\begin{equation}\label{frw}
ds^2 = a(\eta)^2 \left( - d\eta^2 + \gamma_{ij} dx^i dx^j \right)~, \quad i = 1,2,\ldots,d~.
\end{equation}   
Constant conformal time $\eta$ slices are given by three-dimensional slices of constant curvature with induced metric $a(\eta)^2 \gamma_{ij}$. We will restrict our analysis to $\gamma_{ij} = \delta_{ij}$ in what follows. Furthermore we assume periodic boundary conditions ${x}_i \sim {x}_i + L$. The pure de Sitter geometry with flat slicing, which is our main focus, has $a(\eta) =\ell/\eta$ with $\eta \in [-\infty,0]$. Future infinity $\mathcal{I}^+$ resides at $\eta = 0$. The quantity $\ell$ is the de Sitter length which is related to the cosmological constant $\Lambda$ as $\ell^2 \sim 1/\Lambda$. 

We consider the evolution of non-interacting massless scalar fields in this background geometry. Massless fields experience quantum fluctuations which freeze out and reach all the way into the future. This leads to a distribution of possible field configurations on the late time spacelike slice. The massless scalar is the simplest case to consider, and it adequately models the behavior of metric perturbations.

\subsection{Distances in field space} \label{distfieldspace} 

Given two field configurations $\varphi_1(\vec{x})$ and $\varphi_2(\vec{x})$ on a constant time slice, we define a distance between them as follows. First we define new fields $\hat{\varphi}_1$ and $\hat{\varphi}_2$ by subtracting the zeromode and coarse graining over some finite scale by convolution with some window function. The latter is equivalent to imposing a physical size UV cutoff on the field modes. Thus: 
\begin{equation} \label{hatphi}
 \hat{\varphi}(\vec x) = \int d^dy \, w(\vec y) \, \varphi(\vec x+\vec y) - \frac{1}{L^d} \int d^d x \, \varphi(\vec x) \, ,
\end{equation} 
where $w(y)$ is a window function (e.g.\ $w(\vec y) \propto e^{-y^2/\lambda^2}$) which we imagine to have a width of the order of the Hubble radius. The zeromode has no natural normalizable ground state and it is unobservable; subtracting it here will allow us to conveniently discard this mode altogether for our purposes. 

We then define a distance between two $\varphi_1(\vec{x})$ and $\varphi_2(\vec{x})$ as
\begin{equation} \label{distancedef}
d_{12} = d[\varphi_1,\varphi_2] = \frac{1}{L^d} \int d^d x  \left( \hat{\varphi}_1(\vec x)   - \hat{\varphi}_2(\vec x)  \right)^2 ~.
\end{equation}
As it stands, in de Sitter space, the expectation value of $d_{12}$ diverges linearly with proper time in the far future. To avoid this issue, we further subtract off the average $\langle d_{12} \rangle$ with respect to the probablility distribution dictated by the quantum state. Thus we define the following ``renormalized'' distance:
\begin{equation}\label{distance}
{\delta}_{12}  = d_{12}   - \langle d_{12} \rangle~.
\end{equation}
Relatively similar configurations will have negative values of ${\delta}_{12}$ whereas relatively dissimilar vacua will have positive values.

\section{Hartle-Hawking}

In this section we review the construction of the Hartle-Hawking vacuum wave function \cite{PRINT-83-0937 (CAMBRIDGE)}, which we take to be state our massless scalar field is in.

\subsection{Scalar Solutions}

A minimally coupled massless free scalar $\varphi(\eta,\vec{x}) = \sum_{\vec{k}} e^{i \vec{k} \cdot \vec{x} } \varphi_{\vec k}(\eta)$ in an FRW background of the form (\ref{frw}) has action
\begin{equation}\label{action}
 S 
 = \frac{L^d}{2} \sum_{\vec k} \int d \eta \, a^{d-1} \left( \varphi_{\vec k}' \, \varphi_{-\vec k}' - k^2  \, \varphi_{\vec k} \varphi_{-\vec k}\right) \, ,
\end{equation}
where the prime denotes $\eta$ differentiation. The zero mode $\vec k = 0$ drops out of the distance (\ref{distance}), which allows us to discard it altogether. In what follows it is always understood that $\vec k = 0$ is excluded from sums and products. 
The general solution to the equations of motion for $\varphi_{\vec k}(\eta)$ can be written as a linear combination of mode functions:
\begin{equation}\label{solns}
\varphi_{\vec k}(\eta) = A^{+}_{\vec k} v_{k} (\eta) + A_{\vec k}^{-} v_k^*(\eta)~. 
\end{equation}
In Minkowski space $M_{d+1}$ ($a(\eta) \equiv 1$), the canonical mode functions are $v_k(\eta) = \frac{1}{\sqrt{2 k L^d}} e^{i k \eta}$, while in dS$_{d+1}$ ($a(\eta) \equiv -\ell/\eta$) they are
\begin{equation}
 v_k(\eta) = \frac{\sqrt{\pi}}{2}\ell^{-(d-1)/2}\left(-\frac{\eta}{L} \right)^{d/2} H^{(1)}_{d/2} ( k\eta )~,
\end{equation}
where $H^{(1)}(y)$ is the Hankel function of the first kind. Asymptotically $H^{(1)}_n (y) \sim e^{iy}/\sqrt{y}$ when $y\to-\infty$. Specifically in dS$_2$ this becomes $v_k (\eta) = \frac{1}{\sqrt{2 k L}}e^{i k \eta}$, and in dS$_4$ 
\begin{equation}
v_k (\eta) = \frac{1}{\ell(2 k L)^{3/2}} \left( 1 - i k \eta \right) e^{i k \eta}~.
\end{equation}
The normalization is the natural one for canonical quantization but will actually not matter for our purposes.

\subsection{Vacuum state}

The standard Bunch-Davies vacuum state \cite{134659} of a free scalar in dS$_{d+1}$, evolved to a time $\eta=\eta_0$, can be represented as a Gaussian Hartle-Hawking wave functional: 
\begin{equation}
\Psi[\eta_0;\phi] \propto e^{i W[\eta_0;\phi]}~ ,
\end{equation}
where $W$ is a solution to the Hamilton-Jacobi equations, obtained from the action (\ref{action}) evaluated on the complex solutions 
\begin{equation}
 \varphi_{\vec k}(\eta) = \phi_{\vec k} \, \frac{v_k(\eta)}{v_{ k}(\eta_0)}  \, , \qquad \eta \leq \eta_0 \, .
\end{equation} 
This satisfies the boundary conditions $\varphi_{\vec k}(\eta_0) = \phi_{\vec k}$ and $\varphi_{\vec k}(\eta) \sim e^{ik\eta}$ for $\eta \to -\infty$.
Integrating (\ref{action}) by parts reduces the on-shell action to a boundary term and identifies
\begin{equation}
 W[\eta_0;\phi] = \frac{1}{2} a(\eta_0)^{d-1} L^d \sum_{ \vec{k} } \left( \log v_k \right)'(\eta_0) \,  |\phi_{\vec k}|^2~.
\end{equation}
Indeed for this choice the wave functional $e^{i W}$ is a properly damped Gaussian, mimicking for each mode $\vec k$ the canonical Minkwoski vacuum in the far past. 

The probability functional of a given configuration $\phi$ at some time $\eta = \eta_0$ is thus given by 
\begin{equation} \label{PHH}
\mathcal P[\phi]  = | \Psi[\phi] |^2 \propto e^{- 2 \sum_{\vec{k}} \beta_k |\phi_{\vec k}|^2}~, \quad \beta_k \equiv \Re\bigl[ \frac{i}{2} a^{d-1} L^d (\log v_k )' \bigr] \, .
\end{equation}
For example in dS$_2$ we have $\beta_k = k L/2$ and in dS$_4$ we get $\beta_k = \ell^2(L k)^3/2(1+k^2\eta_0^2)$. More generally in the late time limit $\eta_0 \to 0$ we find for dS$_{d+1}$
\begin{equation} \label{betakdS}
 \beta_k = \frac{1}{2} (L k)^d \ell^{d-1} \, .
\end{equation}

\section{Distance Distributions}

We are now ready to compute the distance distributions. We begin by making more precise the notion of  ``branches'' of the wave function of the universe, as states characterized by a definite value for the coarse grained fields. We point out the analogy with thermodynamic pure states in statistical mechanics, in particular spin glass theory. Following this analogy, we define a notion of distance on the state space and introduce their probability distributions. In the theory of spin glasses these serve as order parameters capturing the spin glass phase. We proceed by computing these distributions, first for dS$_2$, for which exact results can be derived, and then for the general dS$_{d+1}$ case. The resulting three point function exhibits ultrametricity.

\subsection{State space and replica reasoning}

Let $\hat{\varphi}$ be the coarse grained field of (\ref{hatphi}). In the Bunch-Davies/Hartle-Hawking vacuum, equal time correlators $\langle \hat{\varphi}(x) \hat{\varphi}(y) \rangle$ are logarithmic and do not vanish at large separation --- in other words the Bunch-Davies vacuum does not satisfy cluster decomposition. Related to this is the fact that $\hat{\varphi}(x)$ does not have a definite value: starting from the Bunch-Davies vacuum, it could freeze into a huge number of possible profiles at late times.

This is analogous to how the usual Boltzmann-Gibbs state does not satisfy cluster decomposition and order parameters do not have definite expectation values whenever there exist different equilibrium states (like the spin up/down states of the Ising model at low temperatures). In analogy to how in spin glass theory one decomposes the Boltzmann-Gibbs state ${\cal P}_{\rm BG} \sim e^{-\beta H}$ into ``pure'' states which do satisfy clustering  \cite{spinglassbook}, we imagine decomposing the Bunch-Davies state into such states. More precisely, we imagine decomposing the Hartle-Hawking probability functional ${\cal P}[\phi]$ (\ref{PHH}) as
\begin{equation}
 {\cal P}[\phi] = \sum_\alpha w_\alpha {\cal P}_\alpha[\phi] \, , \qquad \sum_\alpha w_\alpha = 1 \, .
\end{equation}
Here $\alpha$ is an abstract label for the states exhibiting the clustering property and definite expectation values $\langle \varphi \rangle_\alpha$ for the field $\varphi$. Happily, as in the spin glass case, it is in fact possible to extract detailed information about the structure of the state space without having to perform this decomposition explicitly. The key idea is to consider \emph{probability distributions} of distances or overlaps between pairs of pure states. In the case at hand, the distance between two states $\alpha$ and $\beta$ is 
\begin{equation}
 d_{\alpha \beta} = d[\langle \varphi \rangle_\alpha,\langle \varphi \rangle_\beta] \, ,
\end{equation} 
where $d[\cdot,\cdot]$ is as defined in (\ref{distancedef}) and $\langle \varphi(x) \rangle_\alpha$ is the local vev of $\varphi(x)$ (or more precisely of $\hat{\varphi}(x)$) in the state labeled by $\alpha$. The  probability distribution for finding a distance $D$ between two states is then
\begin{equation}
 P(D) \equiv \sum_{\alpha \beta} w_\alpha w_\beta \, \delta(D-d_{\alpha\beta}) \, , 
\end{equation}
and similarly $P(D_1,D_2,D_3) \equiv \sum_{\alpha \beta \gamma} w_\alpha w_\beta w_\gamma \, \delta(D_1-d_{\alpha\beta}) \, \delta(D_2-d_{\beta\gamma}) \, \delta(D_3-d_{\gamma\alpha})$. As they stand, these formal expressions cannot be evaluated. However, as in the spin glass case, in the thermodynamic limit, one can rewrite this as the distance distributions between \emph{configurations} of a pair of replicas sampled directly from the Bunch-Davies vacuum state:
\begin{equation}
 P(D) = \langle \delta(D-d_{12}) \rangle \equiv \int {\cal D} \phi^{(1)} \, {\cal D}\phi^{(2)} \, |\Psi_{\rm HH}(\phi^{(1)})|^2 \, |\Psi_{\rm HH}(\phi^{(2)})|^2 \,\delta(D-d[\phi^{(1)},\phi^{(2)}]),
\end{equation}
and similarly $P(D_1,D_2,D_3) = \langle \delta(D_1-d_{23}) \, \delta(D_2-d_{31}) \, \delta(D_3-d_{12}) \rangle$, now involving three replicas. As we will see, these distributions are easily computed.

As mentioned already in section \ref{distfieldspace}, the late time expectation value of $d_{12}$ diverges, so we consider the renormalized distances $\delta_{12}$ instead, as defined in (\ref{distance}), with corresponding probability distribution $P(\Delta)=\langle \delta(\Delta-\delta_{12}) \rangle$.



\subsection{Generating function}

We will focus on the distance distribution in the late time limit $\eta_0 \to 0$ of de Sitter space. In the late time limit, the number of Hubble and coarse graining volumes goes to infinity, so this can also be thought of as the thermodynamic limit. In this limit, the UV cutoff on $k$ imposed by coarse graining $\varphi$ is pushed to infinity, so as long as we are computing UV finite quantities, we can remove the cutoff completely, such that only the zeromode subtraction remains in (\ref{hatphi}).

In order to compute $P(\Delta)$ it is convenient to first compute the moment generating function $G(s) = \langle e^{-s \, {\delta}_{12}} \rangle$, and from this obtain $P(\Delta) = \frac{1}{2 \pi i} \int_{-i \infty}^{i \infty} e^{s \Delta} G(s)$. For the probability functional (\ref{PHH}), the function $G(s)$ is given by a simple Gaussian functional integral:
\begin{equation}\label{onepoint}
\langle e^{-s \, {\delta}_{12}} \rangle = e^{\langle s \, \delta_{12}\rangle}   {\prod_{\vec{k}}}' {\cal N}_k
\int d^2\phi_{\vec k}^{(1)} \, d^2\phi_{\vec k}^{(2)} \,
 e^{ -4\beta_k \left|\phi^{(1)}_{\vec k} \right|^2 -4 \beta_k \left|\phi^{(2)}_{\vec k} \right|^2  - 2 s \left|\phi_{\vec k}^{(1)} - \phi_{\vec k}^{(2)} \right|^2}~.
\end{equation}
The primed product runs over unordered pairs $(\vec k,-\vec k)$ with $\vec k \neq 0$, which factorizes the integration over the independent complex variables $\phi_{\vec k} = \phi_{- \vec k}^*$. The resulting Gaussian integrals are proportional to
\begin{equation}
 Z(k,s) \equiv {\det \begin{pmatrix} 2\beta_k + s & -s \\ -s & 2\beta_k + s \end{pmatrix}}^{-1}
\end{equation}
and the proper normalization is obtained by dividing by $\langle 1 \rangle$, i.e.\ ${\cal N}_k = 1/Z(k,0)$. Thus,
\begin{equation} \label{genFgen}
 \langle e^{-s {\delta}_{12}} \rangle = {\prod_{\vec{k}}}' \frac{e^{s/\beta_k}}{1+s/\beta_k} \, ,
\end{equation}
with the factor $e^{s/\beta_k}$ coming from $e^{\langle s \, {w}_{12}\rangle}$.
In a similar fashion, one obtains the moment generating function for the triple distance distribution:
\begin{equation}\label{threepoint}
\langle e^{-s_3 {\delta}_{12}} e^{-s_2 {\delta}_{13}}  e^{-s_1 {\delta}_{23}} \rangle = {\prod_{\vec{k}}}' \frac{{e^{{(s_1+s_2+s_3)}/{\beta_k}}} }{ {1+{(s_1+s_2+s_3)}/{\beta_k}+{3 (s_2 s_3+s_1 s_2+ s_1 s_3)}/{4 \beta_k^2}} }~.
\end{equation}

\subsection{dS$_2$}

\begin{figure}
\begin{center}
\includegraphics[height=6cm]{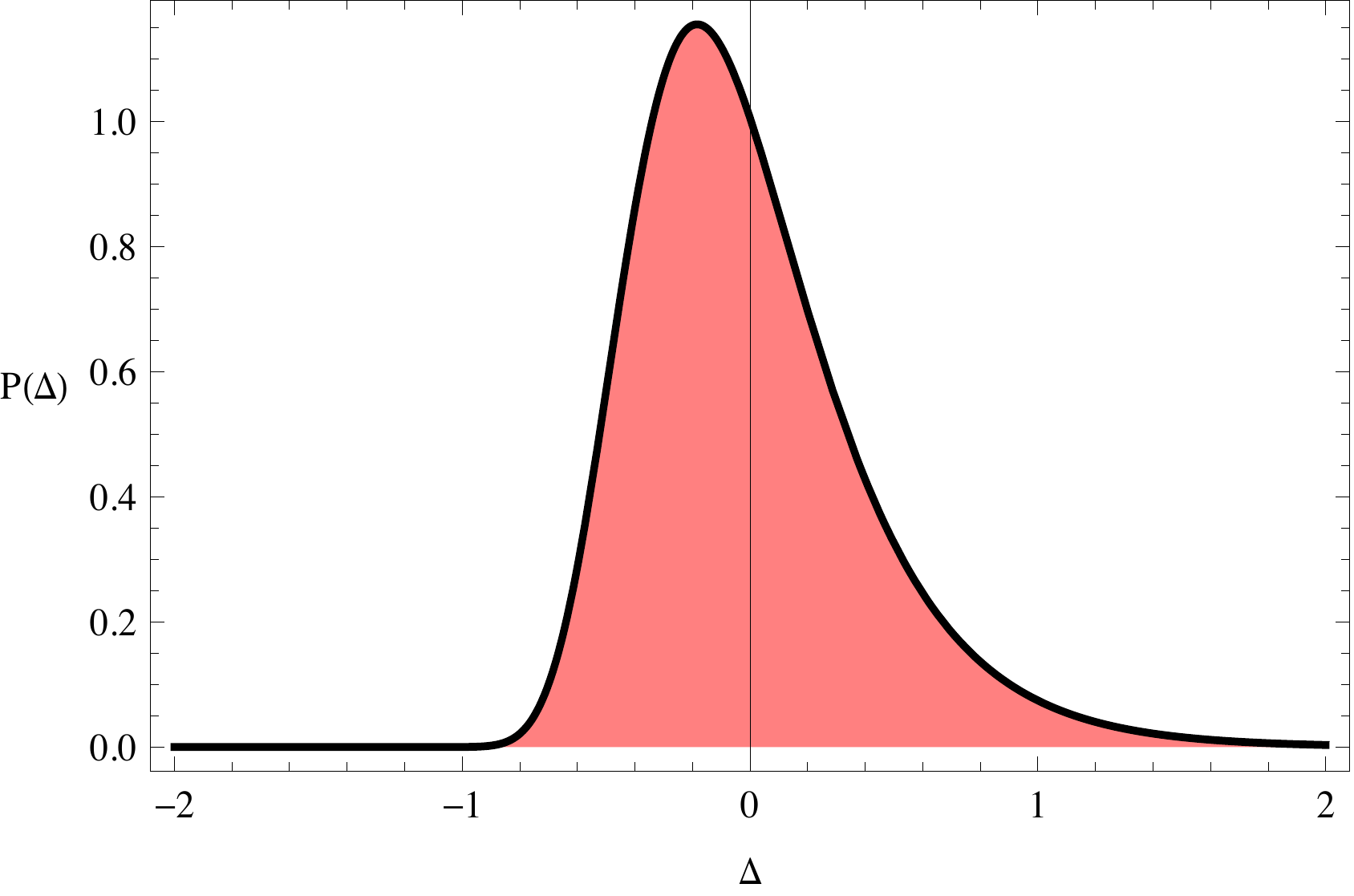}
\end{center}
\caption{Plot of the Gumbel distribution in equation (\ref{gumbelds2}).}\label{gumbelfig}
\end{figure}

In the case of dS$_2$ we have $\beta_k = k L/2 = \pi n$ with $n \in \IZ^+$, and the infinite product (\ref{onepoint}) can be expressed as follows:
\begin{equation}
\langle e^{-s {\delta}_{12}} \rangle = \prod_{ n=1 }^\infty \frac{e^{s/\pi n}}{1+s/\pi n}  = e^{\gamma_E s / \pi} \; \Gamma\left[1+ \frac{s}{\pi} \right]~,
\end{equation}
where $\gamma_E = 0.577216...$ is Euler's $\gamma$ constant. The inverse Laplace transform of the above expression provides $P(\Delta)$. This is easily performed using Euler's integral expression for the $\Gamma$ function given by $\Gamma[z] = \int_0^\infty t^{z-1} e^{-t} dt$. The resulting distribution is a Gumbel distribution:
\begin{equation}\label{gumbelds2}
P(\Delta) = \int^{i\infty}_{-i\infty} \frac{ds}{2\pi i} \, e^{s \Delta }  \; e^{\gamma_E s/\pi} \; \Gamma[1+s/\pi] = \pi \, e^{-(\pi \Delta +\gamma_E) - e^{-(\pi \Delta+\gamma_E)} }~.
\end{equation}
The Gumbel distribution is an extreme value distribution. It appears in the analysis of wide variety of physical systems as discussed e.g.\ in \cite{gumbel}. We depict it in figure \ref{gumbelfig}. 
There is a clear asymmetry: It falls of much faster for negative values than positive ones. This reflects the fact that there are exponentially more dissimilar configurations then there are similar ones. 

\begin{figure}[h!]
\begin{center}
\includegraphics[height=10cm]{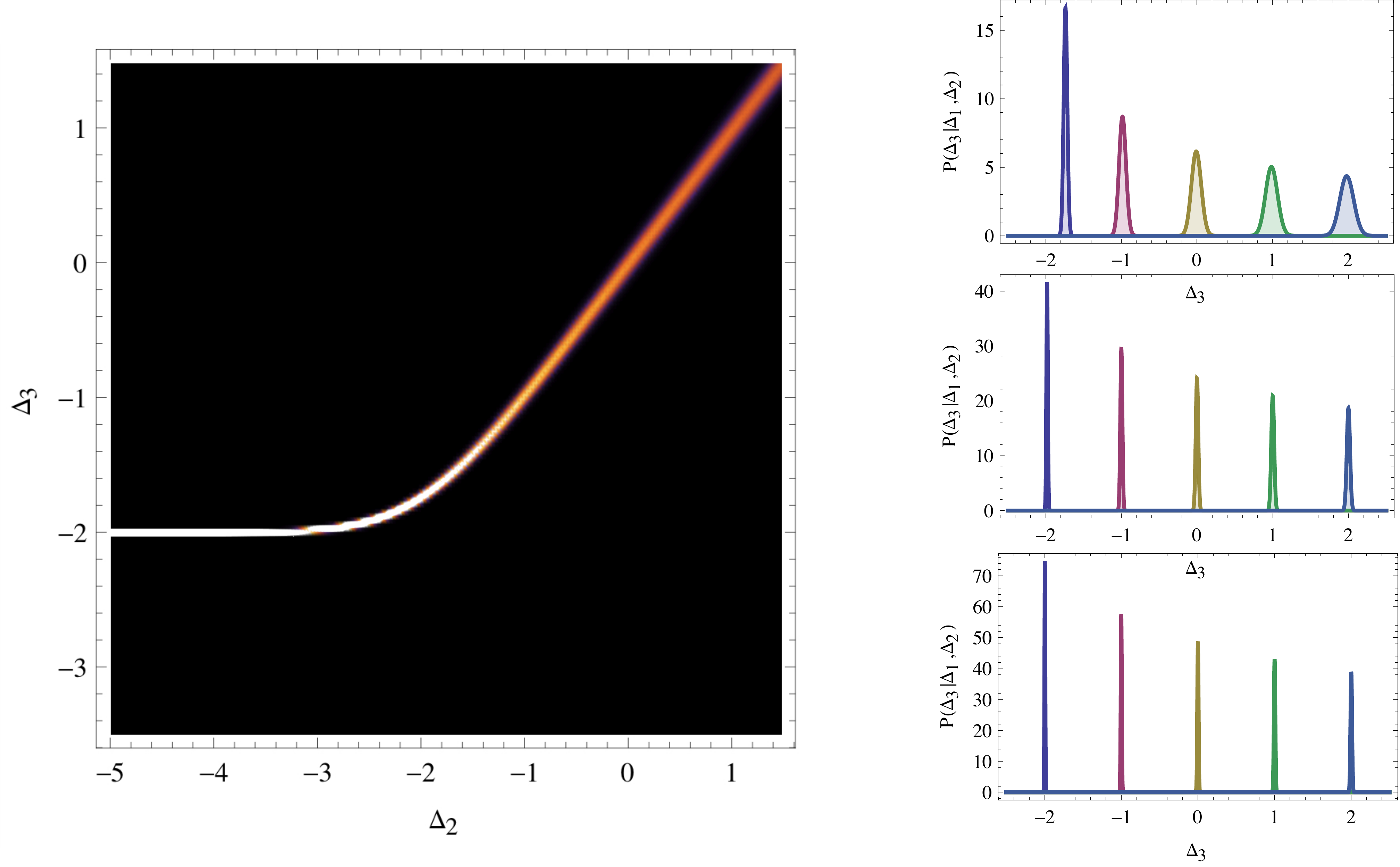}
\end{center}
\caption{Left: Conditional probability $P(\Delta_3|\Delta_1,\Delta_2)$ for $\Delta_1=-2$. Larger values of $P$ correspond to lighter colors. The sharp peaking on $\Delta_3 = \max \{ \Delta_1,\Delta_2 \}$ indicates ultrametricity. Right: $P(\Delta_3|\Delta_1,\Delta_2)$ for  $\Delta_1 = -2,-3,-3.5$ from top panel to bottom panel and $\Delta_2=-2,-1,0,1,2$ from left peak to right peak. The peaks get sharper for lower $\Delta_i$.}\label{ultrametricfig}
\end{figure}

The three point distance distribution can be similarly analyzed in dS$_2$. Once again, the infinite product can be explicitly done. We find:
\begin{equation}\label{triple}
G(\vec s) \equiv \langle e^{-s_3 {\delta}_{12}} e^{-s_2 {\delta}_{13}}  e^{-s_1 {\delta}_{23}} \rangle = e^{\gamma_E L(\vec s) / \pi} \, {\Gamma}\bigl[1 + \tfrac{L(\vec s) + Q(\vec s)}{2 \pi} \bigr] {\Gamma}\bigl[ 1 + \tfrac{L(\vec s)-Q(\vec s)}{2 \pi} \bigr] \, ,
\end{equation}
where we defined the permutation invariants:
\begin{equation}
L(\vec x) \equiv x_1 + x_2 + x_3 \, , \qquad Q(\vec x) \equiv \sqrt{\tfrac{1}{2}(x_1-x_2)^2 + \tfrac{1}{2}(x_2-x_3)^2 + \tfrac{1}{2}(x_3-x_1)^2} ~.
\end{equation}
The three point distribution $P(\vec \Delta)$ is the inverse Laplace transform of (\ref{triple}). We were unable to do this exactly. After replacing the $\Gamma$-functions by their integral representation, a straightforward but somewhat tedious saddle point evaluation yields
\begin{equation} \label{P3qqq}
P(\vec{\Delta}) \propto \exp \left( -\tfrac{2}{3} L(\vec \Delta') - 2 \, e^{-\frac{1}{3} L(\vec \Delta') } \cosh\bigl[\tfrac{2}{3} Q(\vec \Delta')\bigr] \right) ~, 
\end{equation}
where $\vec \Delta'= \pi \vec{\Delta} + \vec \gamma_E$ and $\vec \gamma_E = (\gamma_E,\gamma_E,\gamma_E)$. 

The conditional probability to find the configurations $\phi^{(1)}$ and $\phi^{(2)}$ at a renormalized distance $\Delta_3$ given that $\phi^{(2)}$ and $\phi^{(3)}$ are at a renormalized distance $\Delta_1$ and that $\phi^{(1)}$ and $\phi^{(3)}$ are at a renormalized distance $\Delta_2$ is
\begin{equation}
 P(\Delta_3|\Delta_1,\Delta_2) = \frac{P(\Delta_1,\Delta_2,\Delta_3)}{\int d\Delta_3 \, P(\Delta_1,\Delta_2,\Delta_3)} \, .
\end{equation}
We depict this distribution in fig.\ \ref{ultrametricfig}. A striking structure emerges: the distribution tends to peak very sharply on isosceles triangles --- in other words, the state space is essentially ultrametric!

This can be seen explicitly from (\ref{P3qqq}). First note that $Q(\vec \Delta') \geq \frac{\sqrt{3} \pi}{2} |\Delta_1 - \Delta_2|$, so that if $|\Delta_1 - \Delta_2|$ gets larger than about one, the $\cosh$ function is well approximated by an exponential. The second term in the exponential (\ref{P3qqq}) is then minimized by minimizing $-L(\vec \Delta') + 2Q(\vec \Delta')$. The minimum of this term is $e^{- \pi \min \{\Delta_1,\Delta_2\}}$ and is reached at 
\begin{equation} \label{maxq1q2}
 \Delta_3 = \max \{ \Delta_1,\Delta_2 \} \, ,
\end{equation}
i.e.\ on an isosceles triangle with the unequal side the shortest of the three. When $\min \{\Delta_1,\Delta_2\} < 0$, i.e.\ at least two configurations are closer than average, there will be superexponential suppression of deviations from (\ref{maxq1q2}), and the distribution becomes strongly ultrametric. This is also clearly visible in the figures. Note that in this regime the total probability is always exponentially small as well. This is because we are conditioning on the intrinsically rare event of sampling two relatively similar configurations. The ultrametric structure is for this reason easy to miss if one simply plots the total joint distribution including the region where it gets maximal.

\subsection{dS$_{d+1}$}

For higher dimensional de Sitter, the product (\ref{genFgen}) cannot be computed in closed form.  We can however approximate its logarithm by an integral, taking into account the momentum quantization $k_i \in \frac{2 \pi}{L} \IZ$:
\begin{equation}
 \log \langle e^{-s {\delta}_{12}} \rangle \approx \frac{1}{2} \int d^d k \, \frac{L^d }{(2 \pi)^d} \, \left( \frac{s}{\beta(k)} - \log\left( 1 + \frac{s}{\beta(k)} \right) \right) \, ,
\end{equation}
where for $dS_{d+1}$ in the late time limit $\eta_0 \to 0$, according to (\ref{betakdS}), we have $\beta(k) = \frac{1}{2} (L k)^d \ell^{d-1}$. On the other hand the volume of a shell in momentum space is $\omega_d \, d(k^d)$ where $\omega_d = \frac{\pi^{d/2}}{(d/2)!}$ is the volume of the $d$-dimensional unit ball. Happily this implies the integral takes a universal form for all $d$:
\begin{equation}
 \log \langle e^{-s {\delta}_{12}} \rangle \approx \frac{\omega_d}{(2 \pi)^d \ell^{d-1}} \, \int_{\beta_0}^\infty d\beta \left( \frac{s}{\beta} - \log\left( 1 + \frac{s}{\beta} \right) \right) \, , \qquad \beta_0 \equiv \frac{1}{2} (2 \pi)^d \ell^{d-1} \, .
\end{equation}
The chosen value of the IR cutoff $\beta_0$ corresponds to $k_0 \equiv \frac{2 \pi}{L}$ . This integral is easily computed, and we can evaluate the inverse Laplace transform again in saddle point approximation, yielding again a Gumbel distribution:
\begin{equation}
 P(\Delta) \simeq \frac{\kappa^\kappa}{\Gamma(\kappa)} \, \exp \bigl[ -\kappa ( \Delta'+ e^{- \Delta'}) \bigr] \, , 
\end{equation}
where 
\begin{equation}  \label{kappaDelta}
\kappa \equiv \frac{\omega_d}{2} = \frac{1}{2} \frac{\pi^{d/2}}{(d/2)!} \, , \qquad \Delta' \equiv  \gamma + \frac{\pi}{\kappa} (2 \pi \ell)^{d-1} \Delta \, , \qquad \gamma = \log(\kappa) - \psi(\kappa) \, ,
\end{equation}
with $\psi$ the digamma function. The shift by $\gamma$ and normalization factors do not follow directly from the saddle point computation but are inferred from the requirements $\langle 1 \rangle = 1$, $\langle \Delta \rangle = 0$. In the case $d=1$, we have $\kappa=1$ and $\Delta'=\gamma_E+ \pi \Delta$ so this reproduces the exact result (\ref{gumbelds2}). For $d=3$, we have $\kappa=2\pi/3$ and $\gamma \approx 0.257336$. Notice that the characteristic scale for the distance distribution is set precisely by the dS temperature $T={1}/{2 \pi \ell}$ \cite{125663}.

%

Finally the conditional probability for $\Delta_3$ given $\Delta_1$ and $\Delta_2$ can be obtained by similar approximations. We find:
\begin{equation}
 P(\vec{\Delta}) \propto \exp \left[ -\kappa \left( \tfrac{2}{3} L(\vec \Delta') + 2 \, e^{-\frac{1}{3} L(\vec \Delta') } \cosh\bigl[\tfrac{2}{3} Q(\vec \Delta')\bigr] \right) \right]
\end{equation}
with $\kappa$ and $\Delta'$ as in (\ref{kappaDelta}). This is of the same form as (\ref{P3qqq}). We conclude that up to smoothing effects at a scale set by the de Sitter temperature, the state space of a massless scalar in de Sitter space is universally ultrametric.

\section*{Acknowledgements}

It is a great pleasure to acknowledge T.~Anous, M.~Benna, M.~Douglas, S.~Hartnoll, J.~Maldacena, S.~Shenker and A.~Strominger for valuable discussions. D.A. would also like to thank KU Leuven for its warm hospitality during the completion of this work, and M.~Preciado Lopez for inspiration. This work was supported in part by NSF Grant 0756174, DOE grant DE-FG02-91ER40654 and by a grant of the John Templeton Foundation. The opinions expressed in this publication are those of the authors and do not necessarily reflect the views of the John Templeton Foundation.


\begin{thebibliography}{93}

\bibitem{anniehall} Alvy Singer, ``The Universe is Expanding'', Annie Hall, (1977), \href{http://www.youtube.com/watch?gl=US&v=5U1-OmAICpU}{view}.  


\bibitem{astro-ph/9805201} 
  A.~G.~Riess {\it et al.} [Supernova Search Team Collaboration],
  ``Observational evidence from supernovae for an accelerating universe and a cosmological constant,''
  Astron.\ J.\ \ {\bf 116}, 1009  (1998)
  [astro-ph/9805201].
  
\bibitem{astro-ph/9812133} 
  S.~Perlmutter {\it et al.} [Supernova Cosmology Project Collaboration],
  Astrophys.\ J.\ \ {\bf 517}, 565  (1999)
  [astro-ph/9812133].

\bibitem{SLAC-PUB-2576} 
  A.~H.~Guth,
  ``The Inflationary Universe: A Possible Solution to the Horizon and Flatness Problems,''
  Phys.\ Rev.\ D\ {\bf 23}, 347  (1981).

\bibitem{LEBEDEV-81-229} 
  A.~D.~Linde,
  ``A New Inflationary Universe Scenario: A Possible Solution of the Horizon, Flatness, Homogeneity, Isotropy and Primordial Monopole Problems,''
  Phys.\ Lett.\ B\ {\bf 108}, 389  (1982).
  
\bibitem{183959} 
  A.~A.~Starobinsky,
  ``Dynamics of Phase Transition in the New Inflationary Universe Scenario and Generation of Perturbations,''
  Phys.\ Lett.\ B\ {\bf 117}, 175  (1982).

\bibitem{Denef:2011ee}
  F.~Denef,
  ``TASI lectures on complex structures,''
  [arXiv:1104.0254 [hep-th]].

\bibitem{diothesis}
  D.~Anninos,
  ``Classical and Quantum Symmetries of De Sitter Space,"
  Ph.D Thesis, Harvard University (May 2011)

\bibitem{talks} F.~Denef, talks given at Simons Center Manhattan Seminar, May 2011; Three String Generations at IHES, May 2011; Solvay Workshop on Gauge Theories, Strings and Geometry, May 2011; Holographic Cosmology v2.0 at PI, June 2011.

\bibitem{gr-qc/0111048} 
  S.~Winitzki,
  ``The Eternal fractal in the universe,''
  Phys.\ Rev.\ D\ {\bf 65}, 083506  (2002)
  [gr-qc/0111048].

\bibitem{arXiv:1110.0496} 
  D.~Harlow, S.~Shenker, D.~Stanford and L.~Susskind,
  ``Eternal Symmetree,''
  arXiv:1110.0496 [hep-th].
  
\bibitem{hep-th/0106113} 
  A.~Strominger,
  ``The dS / CFT correspondence,''
  JHEP\ {\bf 0110}, 034  (2001)
  [hep-th/0106113].

\bibitem{hep-th/0106109} 
  E.~Witten,
  ``Quantum gravity in de Sitter space,''
  hep-th/0106109.  
  
\bibitem{astro-ph/0210603} 
  J.~M.~Maldacena,
  ``Non-Gaussian features of primordial fluctuations in single field inflationary models,''
  JHEP\ {\bf 0305}, 013  (2003)
  [astro-ph/0210603].
  
\bibitem{arXiv:1108.5735} 
  D.~Anninos, T.~Hartman and A.~Strominger,
  ``Higher Spin Realization of the dS/CFT Correspondence,''
  arXiv:1108.5735 [hep-th].
  
\bibitem{134659} 
  T.~S.~Bunch and P.~C.~W.~Davies,
  ``Quantum Field Theory in de Sitter Space: Renormalization by Point Splitting,''
  Proc.\ Roy.\ Soc.\ Lond.\ A\ {\bf 360}, 117  (1978).
  
  
\bibitem{PRINT-83-0937 (CAMBRIDGE)} 
  J.~B.~Hartle and S.~W.~Hawking,
  ``Wave Function of the Universe,''
  Phys.\ Rev.\ D\ {\bf 28}, 2960  (1983).

\bibitem{spinglassbook}
  M.~Mezard, G.~Parisi, and M.~A.~Virasoro, \emph{Spin Glass Theory and Beyond.} vol. 9,
  \emph{Lecture Notes in Physics}, (World Scientic, 1987). ISBN 9971501155. \href{http:
  //books.google.com/books?id=ZIF9QgAACAAJ}{Google books}.



\bibitem{LNF-79-31-P} 
  G.~Parisi,
  ``An Infinite Number Of Order Parameters For Spin Glasses,''
  Phys.\ Rev.\ Lett.\ \ {\bf 43}, 1754  (1979).

\bibitem{parisi2} G.~Parisi, ``A sequence of approximated solutions to the SK model for spin glasses,'' J. Phys. A: Math. Gen. 13 L115 (1980).

\bibitem{198330} 
  G.~Parisi,
  ``Order parameter for spin-glasses,''
  Phys.\ Rev.\ Lett.\ \ {\bf 50}, 1946  (1983).

\bibitem{107911} 
  D.~Sherrington and S.~Kirkpatrick,
  ``Solvable Model of a Spin-Glass,''
  Phys.\ Rev.\ Lett.\ \ {\bf 35}, 1792  (1975).

\bibitem{244332} 
  R.~Rammal, G.~Toulouse and M.~A.~Virasoro,
  ``Ultrametricity for physicists,''
  Rev.\ Mod.\ Phys.\ \ {\bf 58}, 765  (1986).

\bibitem{gumbel} 
  T.~Antal, M.~Droz, G.~Gyorgyi, and Z.~Racz,
  ``1/f Noise and extreme value statistics,''
  arXiv:cond-mat/0105599 [hep-th].
  
  
\bibitem{arXiv:1111.4195} 
  M.~K.~Benna,
  ``De (Baby) Sitter Overlaps,''
  arXiv:1111.4195 [hep-th].

\bibitem{125663} 
  G.~W.~Gibbons and S.~W.~Hawking,
  ``Cosmological Event Horizons, Thermodynamics, and Particle Creation,''
  Phys.\ Rev.\ D\ {\bf 15}, 2738  (1977).

%
%



\end{thebibliography}
\end{document}